\documentclass[12pt]{article}
\usepackage{amsmath}
\usepackage{graphicx,psfrag,epsf}
\usepackage{enumerate}
\usepackage{natbib}
\usepackage{amsmath,amsthm,amsfonts,epsfig,amssymb,natbib,eucal,eufrak}
\usepackage{booktabs}
\usepackage{multirow}
\usepackage{siunitx}
\usepackage{qtree}
\usepackage{color}

\usepackage{float}
\floatstyle{plaintop}
\restylefloat{table}

\newcommand{\blind}{0}

\begin{document}

\def\spacingset#1{\renewcommand{\baselinestretch}%
{#1}\small\normalsize} \spacingset{1}


\if0\blind
{
  \title{\bf Follow Up on Detecting Deficiencies: An Optimal Group Testing Algorithm}
  \author{Yaakov Malinovsky\thanks{yaakovm@umbc.edu}\\
    Department of Mathematics and Statistics\\ University of Maryland, Baltimore County, Baltimore, MD 21250, USA}

  \maketitle
} \fi

\if1\blind
{
  \bigskip
  \bigskip
  \bigskip
  \begin{center}
    {\LARGE\bf Follow Up on Detecting Deficiencies: An Optimal Group Testing Algorithm}
\end{center}
  \medskip
} \fi

\bigskip



\spacingset{1.45} 
In a recent volume of Mathematics Magazine (Vol. 90, No. 3, June 2017) there is an interesting article by
Seth Zimmerman, titled {\it Detecting Deficiencies: An Optimal Group
Testing Algorithm}.
A verbatim summary of the article is as follows:
\bigskip

{\it
{Summary.} The use of group testing to locate all instances of disease in a large population of blood samples
was first considered more than 70 years ago. Since then, several procedures have been used to lower the expected
number of tests required. The algorithm presented here, in contrast to previous ones, takes a constructive rather
than a top-down approach. As far as could be verified, it offers the first proven solution to the problem of finding
a predetermined procedure that guarantees the minimum expected number of tests. Computer results strongly
suggest that the algorithm has a Fibonacci-based pattern.
}
\bigskip

The claim in the summary is contradictory to well-known facts reported in the group-testing literature, which
is easily verified, beginning with the work by
\cite{SG1959}, which was cited by S. Zimmerman himself.
Therefore, I feel compelled to offer a number of comments and clarifications.
In addition, I have made some correction of mistaken claim made by \cite{Z2017}.

\begin{enumerate}
\item
The algorithm presented by S. Zimmerman (SZA hereafter) is an improvement
of a dynamic programming (DP) algorithm originally presented
by Milton Sobel and Phyllis A. Groll (\cite{SG1959} pp. 1218-1219), which they called {\it Procedure $R_{3}$}.
The DP structure approach presented by SZA is based on the fact
that the optimal design at stage $t$ (population size $t$) is constructed from
the optimal designs at stages $1,2,\ldots,t-1$. Please see the examples on pages 172--173 \citep{Z2017}.

\item
A nested class of group-testing algorithms \citep{SG1959, S1960, H1976} is defined by the property
that if the positive subset $I$ is identified, the next
subset $I_1$ to be tested is a proper subset of $I$.
By definition, {\it Procedure $R_3$} (and therefore that of SZA) belongs to the nested class of GT algorithms, but has a restriction (assumption (vi) in \cite{Z2017}) and therefore is not optimal in the nested class (with respect to the expected total number of tests).
The optimal nested algorithm, called {\it Procedure $R_{1}$} by \cite{SG1959}, is also a DP algorithm.
To compare the performances of both algorithms, we use the example by \cite{Z2017} on pages 172--173, where
$q=0.9999,\,\,n=6765$, for which
{\it Procedure $R_3$} yields $12.94809$ as the expected number of tests, which is exactly the same as that reported in \cite{Z2017} on page 173.
The result for {\it Procedure $R_1$} is $10.14778$ expected number of tests. It is important to note that the optimal nested procedure $R_1$ is not optimal \citep{S1960} and the optimal procedure for general $n$ is unknown up-to-date.
\item
The original Procedure $R_3$ has a computational complexity proportional to the square of the population size, i.e., $O(n^2)$, where $n$
is the population size.
 The results (Theorems 2 and 3) obtained by \cite {Z2017} allow for the computational complexity
of the DP algorithm to be reduced by at least half.
In addition, S. Zimmerman found
the maximal group size for testing (see page 172 \citep{Z2017}) to be $\displaystyle n_{max}=\left\lceil\frac{log(1-q)}{log(q)} \right\rceil$.
In the above example, this value is equal to $\displaystyle n_{max}=92099$.
Therefore, all it actually allows is an improvement in the the speed of the original {\it Procedure $R_3$.}
It is important to mention that, based on work of \cite{H1976}, the optimal nested procedure $R_1$ has a computational complexity $O(n)$ (without sorting effort).
\item
The claim on page 172 \citep{Z2017} ($q=0.9999$) {\it ``For a population with $n > 6765$, a
test that included more than 6765 samples would always be disadvantageous''} is wrong.
This is easy to check using either procedure $R_3$ or that of SZA, where for  $n\leq 10778$, the optimal approach is to first test all $n$ units.
For example, $n=10,000$ yields $19.20284$ expected number of tests. In contrast, if we follow the suggestion and divide $n=10,000$ into two groups of sizes 6765 and 3235, then the expected total number of tests will be $12.94809+6.34621=19.2943$.
Although the difference is small, the claim is not correct.
\end{enumerate}

{}

\end{document}